# TE-Polarized Surface Plasmon Polaritons (SPPs) in Nonlinear Multi-layer Graphene-Based Waveguides: An Analytical Model

**Mohammad Bagher Heydari** [1,*]

[1,*] School of Electrical Engineering, Iran University of Science and Technology (IUST), Tehran, Iran

[*]Corresponding author: mo_heydari@alumni.iust.ac.ir ; heydari.sharif@gmail.com

**Abstract:** In this article, an analytical model is proposed for the study of Transverse-electric (TE) surface plasmon polaritons (SPPs) in nonlinear multi-layer graphene-based waveguides. Each graphene sheet has been located between two different Kerr-type layers. As special cases of the general, proposed structure, two new nonlinear graphene-based waveguides are introduced and investigated in this paper. The obtained results show that the propagation properties of these exemplary structures are adjustable via chemical potential and nonlinear coefficients. A large value of the effective index, i.e. $n_{eff} = 82$ is obtained for the chemical potential of $\mu_c = 0.15\ eV$ and the nonlinear ratio of $\alpha_2/\alpha_1 = 0.8$ for the second structure at the frequency of 61 THz. The presented study suggests a novel platform in graphene plasmonics, which can be used for the design of innovative THz devices.

## 1. Introduction

Nowadays, graphene has been emerged and known as a new plasmonic material for the design and fabrication of THz devices such as sensors [1, 2], waveguides [3-17], circulator [18, 19], couplers [20, 21], resonators [22, 23], and filters [24, 25]. In the literature, many research articles have been focused on transverse-magnetic (TM) SPP waves for graphene-based structures because these plasmonic waves exhibit higher mode confinement. Compared to TM-polarized SPP waves, TE-polarized SPPs on graphene have a low degree of localization and also only exist in a narrow frequency band [26]. In near-infrared frequencies, metal plasmonics is a very promising candidate for various applications [27, 28].

One way to effectively enhance the field localization of TE-polarized SPP waves is by integrating graphene with nonlinear materials. In the literature, some research articles have investigated TE-polarized SPPs in graphene-based structures containing Kerr-type nonlinear layers [29-31]. For instance, the authors of [30] have studied the propagation features of even and odd modes in a nonlinear graphene-based waveguide. They have shown that the plasmonic properties of their structure are adjustable via chemical potential. In [31], the authors considered the power flow and the field distributions of electromagnetic fields for graphene-based nonlinear structures. It should be noted that TE-polarized SPPs have fascinating properties in multilayer structures [32-34]. The usage of graphene with Kerr-type materials in multilayer structures to excite TE-polarized waves has some advantages: It can enhance the field localization and the nonlinear effect and also stabilize the TE-polarized waves (to prevent the surface waves acting as the air radiation). It can also increase the degrees of freedom for the design of innovative graphene-based devices based on proposed multi-layer structure. However, to the author's knowledge, no comprehensive study has been published to propose an analytical model for the investigation of TE-polarized SPPs in nonlinear multi-layer graphene-based waveguides. In our general multi-layer structure, each graphene sheet has been sandwiched between two different Kerr-type layers. For each layer, we will solve a nonlinear differential equation and then obtain the electromagnetic field components for it.

The remainder of the article is organized as follows. In section 2, the general structure and its analytical model have been introduced and proposed. To derive the field components of TE-polarized SPP waves, a nonlinear differential equation will be solved in this section. Then, two new exemplary structures have been considered and studied as special cases of the general waveguide. The obtained results in section 3 illustrate that integrating graphene with Kerr-type materials enhances the field localization of TE-polarized SPPs and also the propagating features of these waveguides can be tuned by altering the chemical potential and the nonlinear coefficients. Finally, section 4 concludes the article.

## 2. The Proposed Structure and its Analytical Model

The configuration of the proposed structure has been shown in Fig. 1, where a graphene sheet has been placed between two adjacent Kerr-type materials. In this waveguide, we suppose that each nonlinear material in the N-th layer has the permittivity and the permeability of $\varepsilon_N^{NL}, \mu_N^L$, respectively, where the permittivity of Kerr-type nonlinear material is expressed as:

$$\varepsilon_N^{NL} = \varepsilon_N^L + \alpha_N |E|^2 \tag{1}$$

In (1), $\alpha_N$ is the nonlinear coefficient. It should be noted that the indices of *NL, L, and N* denote to nonlinear, linear, and *N*-th layer of the structure. Furthermore, the conductivity of the graphene in the *N*-th layer has well-known relation [35]:

$$\sigma_N(\omega, \mu_{c,N}, \Gamma_N, T) = \frac{-je^2}{4\pi\hbar} Ln\left[\frac{2|\mu_{c,N}| - (\omega - j2\Gamma_N)\hbar}{2|\mu_{c,N}| + (\omega - j2\Gamma_N)\hbar}\right] + \frac{-je^2 K_B T}{\pi\hbar^2(\omega - j2\Gamma_N)}\left[\frac{\mu_{c,N}}{K_B T} + 2Ln\left(1 + e^{-\mu_{c,N}/K_B T}\right)\right] \tag{2}$$

Where $h$ is the reduced Planck's constant, $K_B$ is Boltzmann's constant, $\omega$ is radian frequency, $e$ is the electron charge, $\Gamma_N$ is the phenomenological electron scattering rate for that layer ($\Gamma_N = 1/\tau_N$, where $\tau_N$ is the relaxation time), $T$ is the temperature, and $\mu_{c,N}$ is the chemical potential for the *N*-th layer which can be altered by chemical doping or electrostatic bias [35]. In the general waveguide, as seen in Fig. 1, the thickness of each layer is denoted by $d_N$ ($d_2$ for the second layer, $d_3$ for the third layer and $d_N$ for the N-th layer). Moreover, $\sigma_N$ is the conductivity of graphene layer which is sandwiched between the *N*-th and *N+1*-th layers. To show the difference of various layers, diverse green colors have been utilized in this figure.

Now, we embark to analyze the proposed structure by writing Maxwell's equations inside the Kerr-type nonlinear layers in the frequency domain (suppose $e^{-i\omega t}$):

$$\nabla \times \boldsymbol{E}_N = j\omega\mu_0 \mu_N^L \boldsymbol{H}_N \tag{3}$$

$$\nabla \times \boldsymbol{H}_N = -j\omega\varepsilon_0 \varepsilon_N^{NL} \boldsymbol{E}_N \tag{4}$$

Here, we consider TE-polarized plasmonic waves $(E_y(z), H_x(z), H_z(z))$ propagating in the x-direction ($e^{i\beta x}$) inside the Kerr-type nonlinear layer. By substituting the components of TE waves, i.e. $E_y(z), H_x(z), H_z(z)$, in relations (3)-(4), we obtain the following equation for the N-th layer of Kerr-type nonlinear material:

$$\frac{d^2 E_{y,N}}{dz^2} - \gamma_N^2 E_{y,N} + k_0^2 \mu_N^L \alpha_N E_{y,N}^3 = 0 \tag{5}$$

Where $k_0$ is the free-space wave-number and,

$$\gamma_N^2 = \beta^2 - k_0^2 \varepsilon_N^L \mu_N^L \tag{6}$$

The transverse component of electric fields can be obtained as:

$$H_{x,N} = \frac{-1}{j\omega\mu_0\mu_N^L} \frac{\partial E_{y,N}(z)}{\partial z} \tag{7}$$

$$H_{z,N} = \frac{-\beta}{\omega\mu_0\mu_N^L} E_{y,N}(z) \tag{8}$$

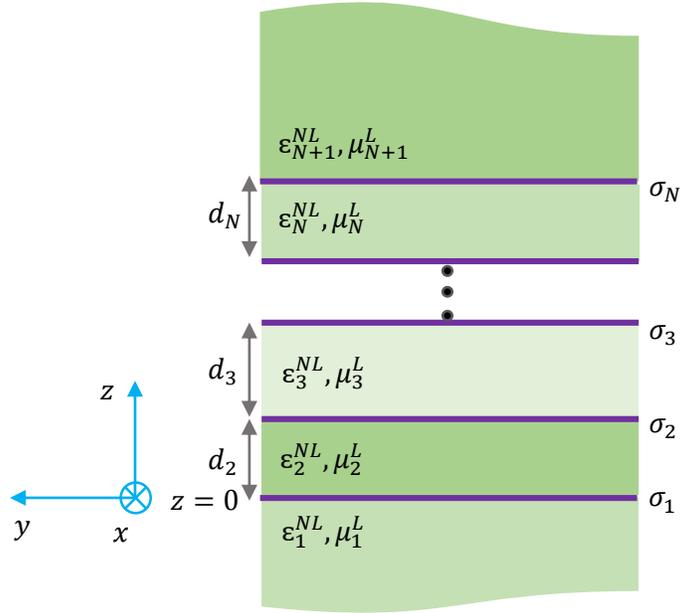

**Fig. 1.** The schematic of the proposed structure. In this waveguide, each graphene sheet has been sandwiched between two different Kerr-type nonlinear layers.

The first integral of (5) results in:

$$\left[\frac{dE_{y,N}}{dz}\right]^2 - \gamma_N^2 E_{y,N}^2 + \frac{1}{2} k_0^2 \mu_N^L \alpha_N E_{y,N}^4 = D_N \tag{9}$$

Where $D_N$ is a constant of integration and,

$$D_N = \left[\frac{dE_{y,N}}{dz}\right]^2 \bigg|_{z=z_N = \sum_{i=2}^{N-1} d_i} - \gamma_N^2 E_{y,N}^2(z_N) + \frac{1}{2} k_0^2 \mu_N^L \alpha_N E_{y,N}^4(z_N) \tag{10}$$

Here, we assume that $D_N \geq 0$ (for the case of $D_N \leq 0$, the expressions are similar). For this case:

$$\frac{d E_{y,N}}{dz} = \pm \sqrt{D_N + \gamma_N^2 E_{y,N}^2 - \frac{1}{2} k_0^2 \mu_N^L \alpha_N E_{y,N}^4} \tag{11}$$

Finally, the solution of (11) is obtained in general form for the N-th layer:

$$E_{y,N}(z) = B_N \, cn\left(q_N\left[(z-z_N) + z_{oc,N}\right], r_N\right) \tag{12}$$

Where following parameters have been defined in (12):

$$q_N = \sqrt[4]{\gamma_N^4 + 2C_N k_0^2 \mu_N^L \alpha_N} \tag{13}$$

$$r_N = \frac{q_N^2 + \gamma_N^2}{2q_N^2} \tag{14}$$

$$B_N = \frac{1}{k_0}\sqrt{\frac{q_N^2 + \gamma_N^2}{\alpha_N}} \tag{15}$$

In (12), "$cn$" is Jacobi elliptic function and $r_N$ is Jacobi modulus. One of the special cases occurs for $D_N = 0, r_N = 1$, which results in

$$E_{y,N}(z) = \frac{\gamma_N}{k_0}\sqrt{\frac{2}{\alpha_N}} \, sech\left(\gamma_N\left[(z-z_N) + z_{oc,N}\right]\right) \tag{16}$$

The final step is applying the boundary conditions for a graphene sheet sandwiched between two Kerr-type nonlinear materials:

$$E_{y,N+1} = E_{y,N} \tag{17}$$

$$H_{x,N+1} - H_{x,N} = \sigma_N E_{y,N} \tag{18}$$

Now, achieving the dispersion relation and other propagating properties is straightforward. In the appendix, we will show how the above boundary conditions can be used for a particular structure (as a special case of the general waveguide of Fig. 1) to obtain a dispersion relation for that case.

## 3. Special Cases of the Proposed Structure: Results and Discussions

This section investigates two new waveguides as special cases of the general, proposed structure (see Fig. 1). The first one is constructed of a nonlinear slab layer deposited on $SiO_2$-Si layers and a graphene sheet is placed on the top surface of the nonlinear slab. The second waveguide is a hybrid nonlinear structure, constituting Nonlinear-Graphene-$SiO_2$-Si-$SiO_2$-Graphene-Nonlinear layers. It will be shown that the usage of two nonlinear layers enhances the propagation features of TE-polarized SPP waves.

In these exemplary structures, graphene sheets have a thickness of 0.33 nm, are at the temperature of 300 K and their chemical potential is 0.15 eV, unless otherwise stated. To excite the TE-polarized SPP waves, the frequency should be in the range of $1.67 < \hbar\omega/\mu_c < 2$ [26]. Thus, by supposing $\mu_c = 0.15 \, eV$, the frequency range will be in the range of $60.5 \, THz < f < 72.5 \, THz$. The permittivities of $SiO_2$ and Si layers are 2.09 and 11.9, respectively. The geometrical parameters in all structures have been considered $t = 0.25 \mu m, d = 0.6 \mu m, s = 0.7 \mu m$. For all nonlinear layers in these structures, the linear permittivity (See relation (1)) is 2.89, i.e. $\varepsilon_L = 2.89$. The electric field intensity is $E_0 = 10^6 \, V/m$. For the first structure, the nonlinear Kerr coefficient is $\alpha|E_{y,0}|^2 = 0.02$ and for the second one, the nonlinearity ratio is $\alpha_2/\alpha_1 = 0.2$, unless otherwise stated. These parameters have been chosen for the frequency range above 1 THz for the electric field intensity of $E_0 = 10^6 \, V/m$, which have been given and explained in the literature [36-44]. In what follows, we will discuss the analytical results provided by MATLAB for these exemplary waveguides.

*3.1 The first structure: A nonlinear slab waveguide incorporating a graphene sheet*

In Fig. 2, the configuration of the first structure is shown, where a graphene sheet has been located on nonlinear -SiO$_2$-Si layers. Fig. 3 represents the propagation properties of the first waveguide as a function of frequency for various values of incident mode power. In Fig. 3(a), the effective index is defined as $n_{eff} = Re[\beta]/k_0$ and the propagation length has been considered as $L_{Prop} = 1/Im[\beta]$ in Fig. 3(b). As seen in Fig. 3, the effective index decreases as the frequency increases while the propagation length increases. Moreover, as the incident mode power increases, the effective index increases but the propagation length decreases at a specific frequency. Indeed, the propagation performance can be enhanced by altering the nonlinear coefficient. For instance, consider $f = 66.5 \, THz$. In this frequency, the effective index reaches to $n_{eff} = 39$ for $\alpha|E_{y,0}|^2 = 0.08$ while it is $n_{eff} = 18$ for $\alpha|E_{y,0}|^2 = 0.02$.

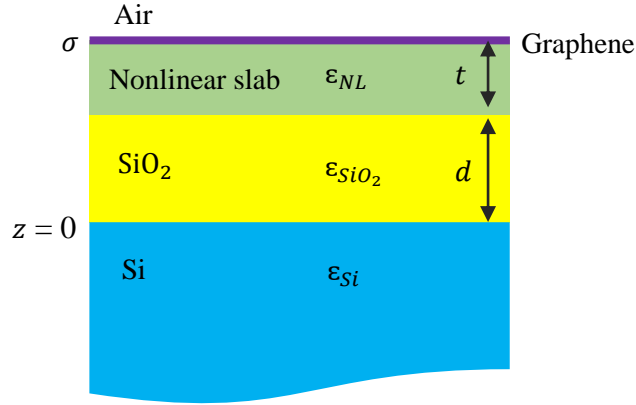

**Fig. 2.** The schematic of the first structure: a graphene sheet has been located on Nonlinear-SiO$_2$-Si layers.

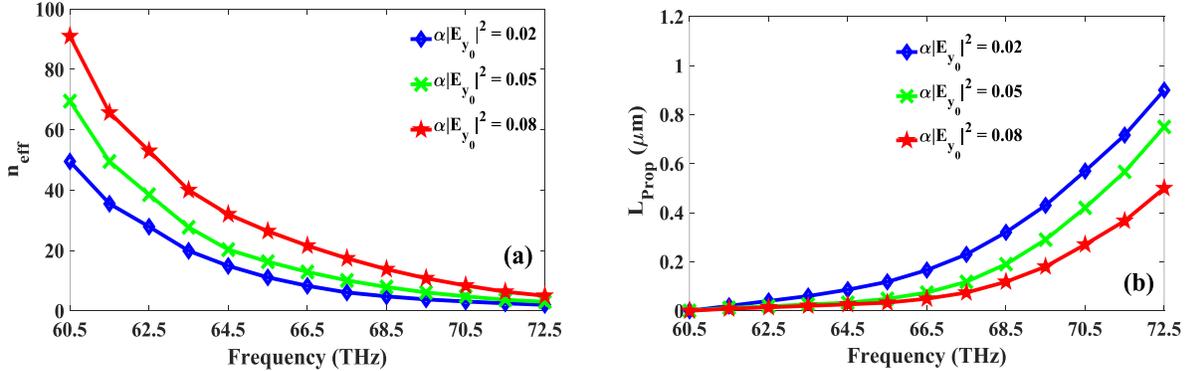

**Fig. 3.** The propagating properties of the first structure as a function of the frequency for various values of the incident mode power.

One of the important parameters for controlling the plasmonic features in graphene-based structures is altering the chemical potential of graphene. Fig. 4 demonstrates the variations of the effective index and the propagation length as a function of the chemical potential at two different frequencies. It can be observed that the effective index decreases with the increment of the chemical potential while the propagation length increases. Furthermore, the slope of variations gets slowly for the potential range of $\mu_c \geq 0.6\ eV$.

As a final point for this sub-section, let us investigate the influence of the incident mode power on the plasmonic features of the first waveguide. As seen in Fig. 5(a), the effective index increases as the value of the nonlinear coefficient increases. However, the propagation length decreases with the increment of the incident mode power, as shown in Fig. 5 (b). Furthermore, the slope of the effective index variations is high for the frequency of $f = 66.5\ THz$. Consider the incident mode power of $|E_{y,0}|^2 = 0.8$. For this nonlinear value, the effective index reaches 82 at the frequency of 66.5 THz while it reaches to $n_{eff} = 21$ at $f = 72.5\ THz$. The propagation length for these frequencies at the incident mode power of $|E_{y,0}|^2 = 0.6$ are 0.18 and 0.16μm, respectively. Hence, the adjustment and controllability of SPPs in this structure can be utilized in nonlinear graphene plasmonics.

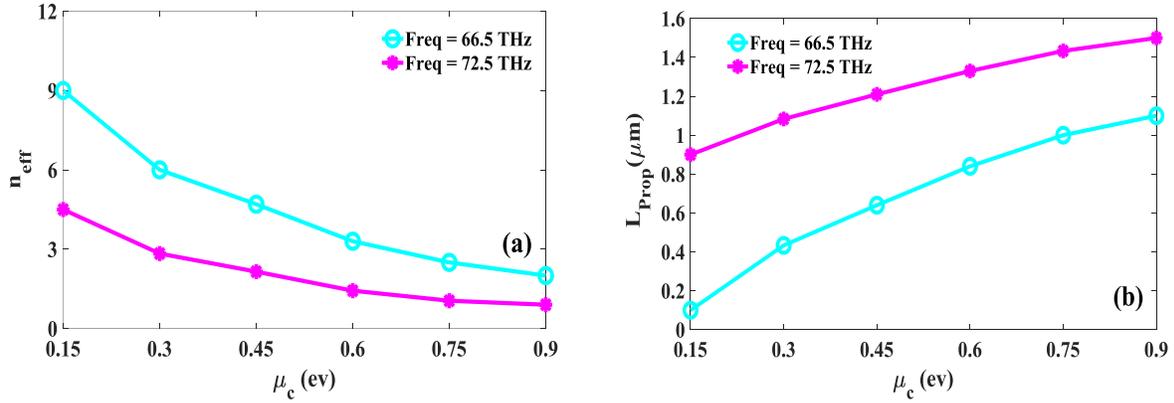

**Fig. 4.** The analytical results of **(a)** the effective index and **(b)** the propagation length, for the first structure as a function of the chemical potential at various frequencies.

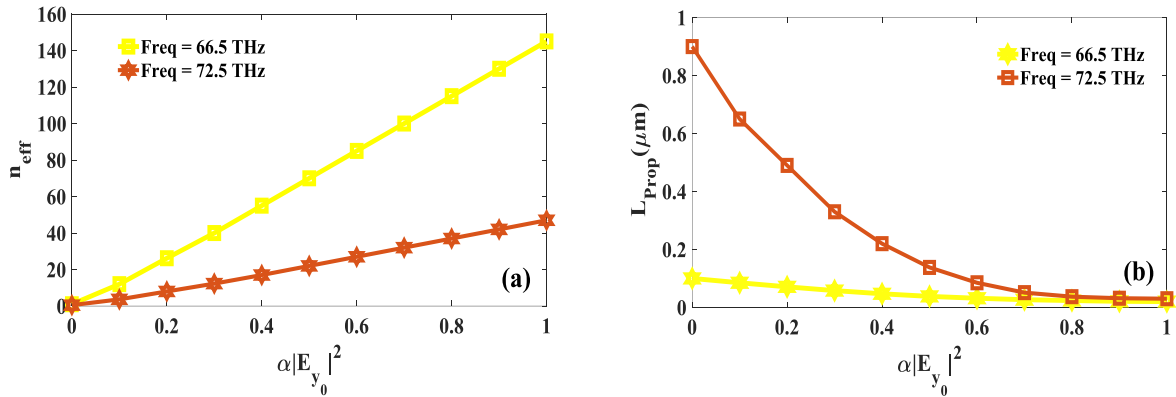

**Fig. 5.** The analytical results of **(a)** the effective index and **(b)** the propagation length, for the first structure as a function of the incident mode power at various frequencies.

*3.2 The second structure: A hybrid graphene-based waveguide with two nonlinear layers*

As a second waveguide, we introduce and study a hybrid nonlinear graphene-based waveguide containing two nonlinear Kerr-type layers. As observed in Fig. 6, this structure is structurally symmetric but the nonlinear parameters for the first and second Kerr layers are different. In Fig. 7, the propagation properties have been depicted for this waveguide as a function of frequency. It is clear from this figure that the effective index decreases as the frequency increases. Moreover, by altering the nonlinearity ratio ($\alpha_2/\alpha_1$), the propagation features change. At a specific frequency, the increment of the nonlinear ratio results in the increment of the effective index. Another important point concluded from this figure is unremarkable variations of effective index and propagation length as the nonlinear ratio varies. This occurs due to the small difference between $\alpha_1$ and $\alpha_2$ ($\alpha_2/\alpha_1 < 1$).

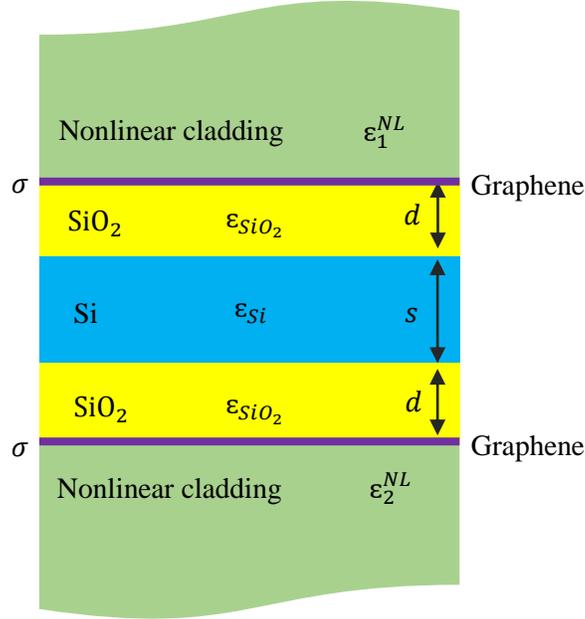

**Fig. 6.** The schematic of the second structure: a hybrid nonlinear waveguide, where two nonlinear layers at both sides of the whole waveguide have been utilized as cladding layers.

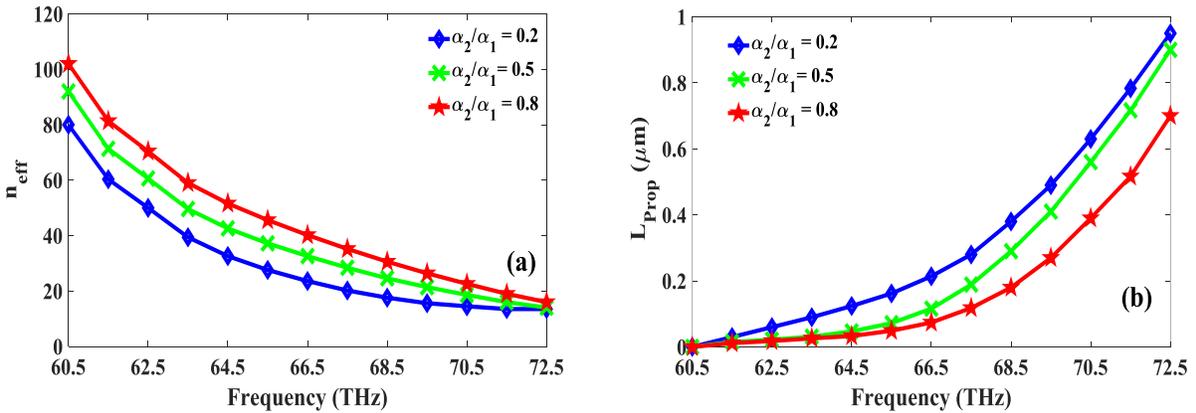

**Fig. 7.** The propagating properties of the second structure as a function of the frequency for various values of the incident mode ratio.

To further consider the influence of chemical potential variation on the plasmonic features, we have depicted Fig. 8 for the second waveguide at two specific frequencies. As observed in this figure, the slope variation of the effective index for the frequency of $f = 72.5\ THz$ is so slow. Furthermore, it is seen that at a specific chemical potential, the higher effective index (and lower propagation length) is obtained at the lower frequency, which confirms the analytical results of Fig. 7.

As a final point, the variations of propagation properties have been demonstrated as a function of the nonlinearity ratio in Fig. 9. In this diagram, it is assumed that the nonlinear Kerr coefficient for the first layer (i.e. $\alpha_1|E_{y,0}|^2 = 0.02$) is constant and the nonlinear coefficient of the second layer (i.e. $\alpha_2$) varies. As the nonlinearity of the second layer increases, the effective index reaches high values at the middle frequency of the studied band ($f = 66.5\ THz$). At the frequency of $f = 72.5\ THz$, the slope of effective index variations is not noticeable because this frequency is the end of the frequency window ($60.5\ THz < f < 72.5\ THz$) and the nonlinearity disappears at this frequency.

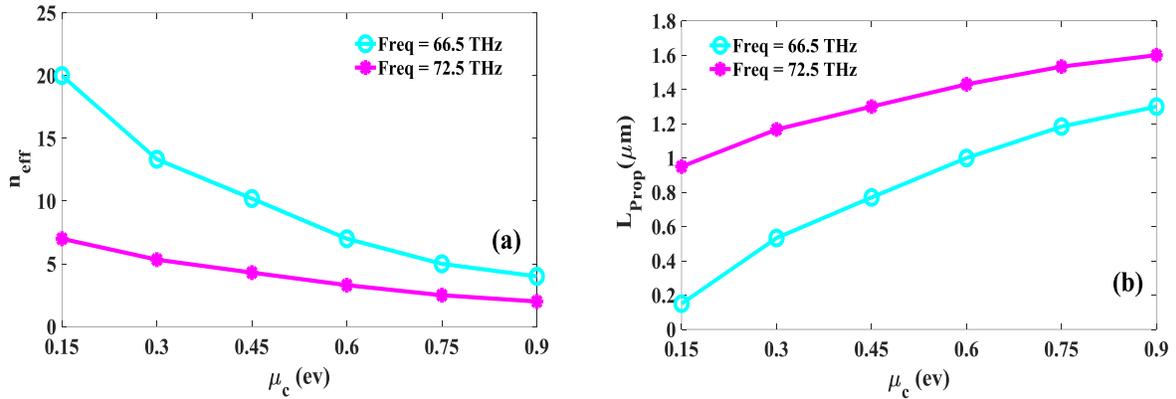

**Fig. 8.** The analytical results of **(a)** the effective index and **(b)** the propagation length, for the second structure as a function of the chemical potential at various frequencies.

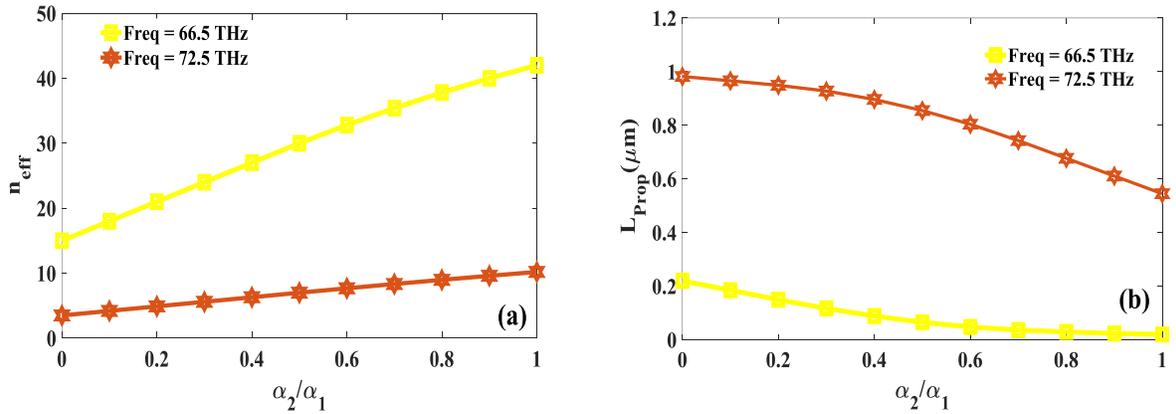

**Fig. 9.** The analytical results of **(a)** the effective index and **(b)** the propagation length, for the second structure as a function of the incident mode ratio at various frequencies.

## 4. Conclusion

This article presented an analytical study of TE-polarized SPPs in nonlinear multi-layer graphene-based waveguides. To illustrate the richness of the analytical model and proposed general structure, two novel waveguides were investigated. The first one was a nonlinear slab waveguide incorporating a graphene sheet. A large value of the effective index ($n_{eff} = 65$) at the frequency of 61 THz was reported for this waveguide for $\mu_c = 0.15\ eV, \alpha|E_{y,0}|^2 = 0.08$. The second structure was a hybrid graphene-based waveguide with two nonlinear Kerr-type layers. Harnessing two nonlinear layers and two graphene sheets resulted in a higher value of propagation properties. For two exemplary waveguides, it was shown that plasmonic features can be adjusted by varying the chemical potential and nonlinear coefficients. Our presented analytical model can be utilized for the design of novel graphene-based components in the THz region. Moreover, the integration of graphene and nonlinear Kerr-type materials enhances the localization of TE-polarized SPP waves.

**Appendix**

In this appendix, we will show how a dispersion relation for a special case of the general structure can be obtained by substituting the relations (12) and (16) into the boundary conditions of (17)-(18). Consider Fig. 10, where a graphene sheet is placed between nonlinear and linear layers. By using the relations of (16) and (7), the electromagnetic components inside each layer can be written as follows:

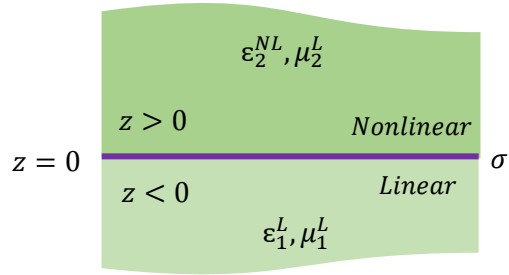

**Fig. 10.** The schematic of the special case of the general structure. In this waveguide, each graphene sheet has been sandwiched between a nonlinear and a linear layer.

$$E_y(z) = \begin{cases} \dfrac{\gamma_2}{k_0}\sqrt{\dfrac{2}{\alpha_2}}\dfrac{1}{\cosh(\gamma_2(z-z_2))} & z > 0 \\ E_0\exp(\gamma_1 z) & z < 0 \end{cases} \quad (19)$$

$$H_x(z) = \dfrac{-1}{j\omega\mu_0}\begin{cases} \dfrac{\gamma_2^2}{k_0}\sqrt{\dfrac{2}{\alpha_2}}\dfrac{\sinh(\gamma_2(z-z_2))}{[\cosh(\gamma_2(z-z_2))]^2} & z > 0 \\ E_0\gamma_1\exp(\gamma_1 z) & z < 0 \end{cases} \quad (20)$$

By applying (17), the second integral constant ($z_2$) is obtained:

$$z_2 = \dfrac{1}{\gamma_2}\cosh^{-1}\left[\dfrac{\gamma_2}{E_0 k_0}\sqrt{\dfrac{2}{\alpha_2}}\right] \quad (21)$$

Now, by substituting (20) into the relation (18) and doing some mathematical procedures, and utilizing (21), we finally archive to the dispersion relation:

$$\gamma_2 \tanh(\gamma_2 z_2) + \gamma_1 = j\omega\mu_0\sigma(\omega) \tag{22}$$

**Declarations**

**Ethics Approval:** Not Applicable.

**Consent to Participate:** Not Applicable.

**Consent to Publish:** Not Applicable.

**Funding:** The author received no specific funding for this work.

**Competing Interests:** The author declares no competing interests.

**Availability of Data and Materials:** Not Applicable.